\documentclass[11pt,a4paper]{article}
\pdfoutput=1

\usepackage{amssymb,amsmath,euscript,array}
\usepackage{subfigure}
\usepackage{color}
\usepackage{cite}
\usepackage{footnote}
\usepackage{caption}
\usepackage{hyperref}
\usepackage{epsfig,graphicx}
\usepackage{rotating}
\usepackage{cancel}
\usepackage{bm}
\usepackage{comment}

\def\beq{\begin{equation}}
\def\eeq{\end{equation}}

\begin{document}
\numberwithin{equation}{section}
\title{{\normalsize  DCPT/11/148; IPPP/11/74 \hfill\mbox{}\hfill\mbox{}}\\
\vspace{2.5cm} \LARGE{\textbf{Direct SUSY Searches at the LHC\\ in the light of LEP Higgs Bounds
\vspace{0.5cm}}}}

\author{\textbf{David Grellscheid, Joerg Jaeckel, Valentin V. Khoze, Peter Richardson} \\[0.3cm] \textbf{and Chris Wymant}\\[2ex]
\small{\em Institute for Particle Physics Phenomenology} \\
\small{\em Department of Physics, Durham University} \\
\small{\em Durham DH1 3LE, United Kingdom}\\[0.5ex]
}
\date{}
\maketitle

\begin{abstract}
\noindent
In this note we compare the latest $1.04\,{\rm fb}^{-1}$ LHC searches for squarks and gluinos from jets and missing transverse momentum (MET)
with constraints arising from the LEP Higgs bound. For General Gauge Mediation models with moderate values of
$\tan (\beta)$ we find that the zero-lepton MET searches of supersymmetry at the LHC are only starting to be competitive
with the Higgs bounds from LEP. From this perspective and for such models, the SUSY searches at the LHC
are still very much in the beginning.
\end{abstract}

\newpage

\section{Introduction}

The LHC has completed its first full year of collecting data and searching for new physics in 7 TeV proton-proton collisions.
The ATLAS and CMS collaborations conducted an impressive series of searches for supersymmetry (SUSY)
at the LHC looking for squarks and gluinos in final states containing jets and missing
transverse momentum~\cite{ATLAS-summary,CMS-summary,ATLAS-0lept,CMS-0lept}. The latest of these analyses are based on $\sim 1\,{\rm fb}^{-1}$ of data taken
in the first half of 2011~\cite{ATLAS-summary,CMS-summary,ATLAS-0lept,CMS-0lept}.

So far, these searches have shown no evidence for SUSY (nor for any other new physics). This fact by itself
does not imply the demise of supersymmetry, it rather points out that new physics was not `hiding around the corner'
and more work is needed to uncover it.
In this paper we will assess the impact of these searches
for General Gauge Mediated (GGM) SUSY breaking models with long-lived next-to-lightest supersymmetric particles (NLSPs).
Interpretations of the latest LHC data in terms of other phenomenological models are given in ~\cite{Allanach:2011qr,Buchmueller:2011sw,Essig:2011qg,Kats:2011qh,Brust:2011tb,Papucci:2011wy}.
Here, we will derive
the exclusion contours arising from the ATLAS zero-lepton search for SUSY at $1.04\,{\rm fb}^{-1}$ \cite{ATLAS-0lept}, henceforth `jets + MET', and compare with the LEP Higgs
bounds~\cite{Schael:2006cr}.
We will argue that for GGM models with low-to-moderate values of $\tan(\beta)$,
the Higgs bounds are still the strongest bounds in sizable areas of parameter space. As such for this classes of models supersymmetry
searches at the LHC are only beginning.

\section{Main Results: Summary and Interpretation}

\begin{figure}
 \hspace*{5cm}\includegraphics[width=4.5cm]{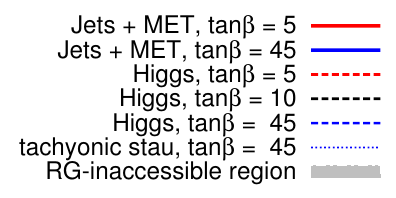}
\\
\hspace*{-1.1cm}
  \includegraphics[width=8cm]{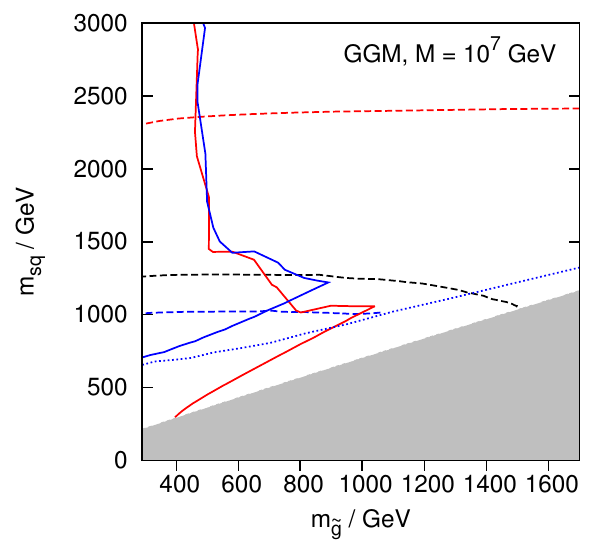}
  \hspace*{0.3cm}
  \includegraphics[width=8cm]{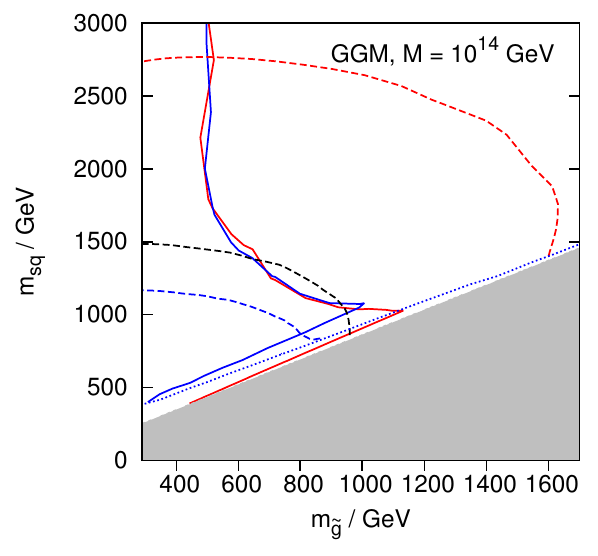}
\begin{center}
\caption{Exclusion contours for General Gauge Mediation from the ATLAS $1.04\,{\rm fb}^{-1}$ jets + MET SUSY search versus
Higgs bounds from LEP in terms of physical squark and gluino masses. The left panel shows GGM with a messenger scale $M=10^7$ GeV, the
right panel $M=10^{14}$ GeV. Jets + MET exclusions are shown as solid lines
with $\tan (\beta) = 5$ in red and $\tan (\beta) = 45$ in blue. Higgs exclusion contours obtained from {\tt HiggsBounds}
are shown as dashed lines in red, black and blue, for $\tan (\beta) = 5$, $10$ and $45$
respectively. For $\tan(\beta)=45$ there is a region where the stau becomes tachyonic, indicated by the dotted blue line.
The grey area is theoretically inaccessible~\cite{Jaeckel:2011wp}.}
\label{fig:GGM-sq-gl-plane}
\end{center}
\end{figure}

Figure~\ref{fig:GGM-sq-gl-plane} summarises our main findings for the LHC SUSY and LEP Higgs exclusion contours
in models of General Gauge Mediation\footnote{We refer the reader to Sect.~\ref{ggm} and references therein for details on GGM.} in the context of the minimal supersymmetric standard model. We plot both sets of exclusions on the plane of physical squark (vertical axis) and
gluino masses (horizontal axis).
The left panel shows GGM with lower messenger masses of $M=10^7$ GeV, and the
right panel shows a high-messenger-mass case with $M=10^{14}$ GeV.
The SUSY exclusion contours in GGM are derived from $1.04\,{\rm fb}^{-1}$ of data collected
by the ATLAS collaboration in their searches for jets and missing transverse momentum (MET) in proton-proton collisions at 7 TeV
with zero leptons in the final state \cite{ATLAS-0lept}.
We obtain these SUSY exclusion contours following the same strategy as in \cite{Dolan:2011ie}.
As will be explained in more detail in the following section, we use
\textsf{Herwig++}~\cite{Bahr:2008pv,Gieseke:2011na} to simulate events in GGM and \textsf{RIVET}~\cite{Buckley:2010ar}
to implement the experimental cuts used by ATLAS in \cite{ATLAS-0lept}. The resulting signal rates of GGM are then compared to
data.

The data strongly constrains events with missing energy carried for example by a neutralino NLSP.
In sizable regions of GGM parameter space the NLSP is the stau (rather than the neutralino), which we
take to be stable on collider scales.  In our analysis we use a conservative approach to the data by assuming that events with long-lived stau NLSPs will not give MET signatures and thus will not be constrained. In these regions one should instead use the dedicated searches for long-lived charged particles. These searches have already started~\cite{ATLAS-stau,CMS-stau} and we plan to return to these analysis in forthcoming work.

The red and blue solid lines in each panel of Fig.~\ref{fig:GGM-sq-gl-plane} denote these MET exclusions in GGM for low
and for high values of $\tan (\beta)$, with $\tan (\beta) = 5$ blue and $\tan (\beta) = 45$ given by the red curve.
Following the contour from the upper left, it can be easily seen
that the bounds from jets + MET searches are largely $\tan (\beta)$-insensitive until one reaches the turning point where the NLSP changes from the lightest neutralino to the stau. This point itself is $\tan(\beta)$ dependent  as can be seen from the figure. Beyond this point the curve follows the boundary between neutralino
and stau NLSP regions. As can be seen from Fig.~\ref{fig:GGM-sq-gl-plane} the stau NLSP region grows at large $\tan(\beta)$ because the stau mass is reduced by its Yukawa coupling, which increases
with $\tan(\beta)$.

GGM models also allow for NLSPs decaying inside the detector. This can be either prompt or with a displaced vertex.
This case has been studied, e.g. in~\cite{Kats:2011qh}.
An example is a bino-like
neutralino decaying into a photon and gravitino. In this case it is more appropriate to rely on the dedicated MET searches with two photons
in the final state \cite{ATLAS-gamma2,CMS-gamma2}. These are more sensitive than the inclusive
jets + MET searches, since the missing transverse momentum is halved when each neutralino can decay into a photon and a gravitino
(with only the latter carrying the MET). In the present paper we concentrate on long-lived NLSPs.

\newpage
The Higgs-based exclusion contours for GGM\footnote{Quite clearly, to talk about Higgs exclusions contours one has to use non-trivial information
about the SUSY spectra. This data is usually not contained in simplified models. Therefore it is desirable to use complete models, as we do here.}  are obtained by using FeynHiggs~\cite{Frank:2006yh,Degrassi:2002fi,Heinemeyer:1998np,Heinemeyer:1998yj} and {\tt HiggsBounds}~\cite{Bechtle:2008jh,Bechtle:2011sb} (for details see the next section).
This automatically enforces the most-constraining-to-date bound on the Higgs sector of each model. In our case this reduces to the LEP searches for SM-like Higgs bosons, so we will refer to these exclusions collectively as the LEP Higgs bounds.

\begin{figure}
 \hspace*{1.2cm}
 \includegraphics[width=4.5cm]{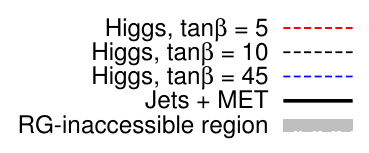}
 \hspace*{4cm}
 \includegraphics[width=4.5cm]{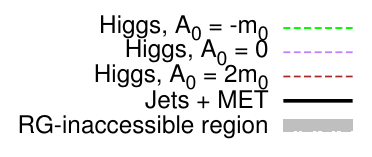}
\\
\hspace*{-1.1cm}
  \includegraphics[width=8cm]{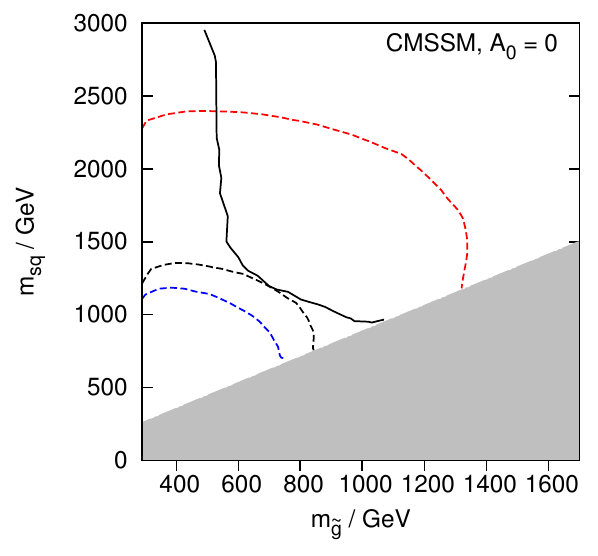}
  \hspace*{0.3cm}
  \includegraphics[width=8cm]{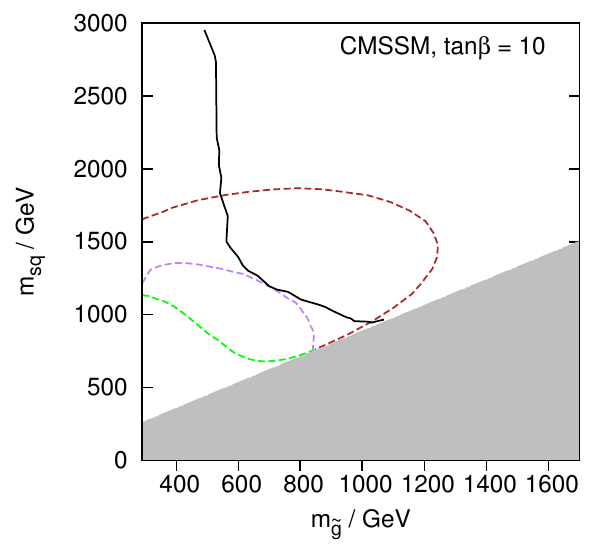}
\begin{center}
\caption{Comparison of the ATLAS $1.04\,{\rm fb}^{-1}$ jets + MET exclusion (solid line) with the LEP Higgs bounds (dashed lines) for the CMSSM. In the left panel we set $A_0=0$ and show the
Higgs bounds for $\tan(\beta)=5$ in red, $\tan(\beta)=10$ in black and $\tan(\beta)=45$ in blue.
On the right panel we fix $\tan(\beta)=10$ and show $A_0 = -m_0, 0, +2 m_0$ in green, purple and brown.}
\label{fig:CMSSM-sq-gl-plane}
\end{center}
\end{figure}

The LEP Higgs bounds for GGM are shown
in Fig.~\ref{fig:GGM-sq-gl-plane} as dashed lines
in red, black and blue, for $\tan (\beta) = 5$, $10$ and $45$ respectively.
The position of the Higgs exclusion contours in Fig.~\ref{fig:GGM-sq-gl-plane} shows significant dependence on the value
of $\tan (\beta)$. This is as expected, since the tree level contribution to the mass of the lightest Higgs is
$m^{\rm tree}_{h^0} = m_Z \cos(2\beta)$, which decreases with decreasing $\tan (\beta)$, making it harder to satisfy the LEP lower
bound on the Higgs mass.
In order to fulfill this bound, as is well known, one needs large loop corrections from the stop.
At one-loop level (see e.g. \cite{Primer}) one has,
\begin{eqnarray}
m^2_{h^0} \>=\> m_Z^2 \cos^2(2\beta) +
\frac{3 }{4 \pi^2} \sin^2(\beta) \>y_t^2 \biggl [
m_t^2 \, {\rm ln}\left (m_{\tilde t_1} m_{\tilde t_2} / m_t^2 \right )
+ c_{\tilde t}^2 s_{\tilde t}^2 (m_{\tilde t_2}^2 - m_{\tilde t_1}^2)
\, {\rm ln}(m_{\tilde t_2}^2/m_{\tilde t_1}^2)
\nonumber
\\
+ c_{\tilde t}^4 s_{\tilde t}^4 \Bigl \lbrace
(m_{\tilde t_2}^2 - m_{\tilde t_1}^2)^2 - \frac{1}{2}
(m_{\tilde t_2}^4 - m_{\tilde t_1}^4)
\, {\rm ln}(m_{\tilde t_2}^2/m_{\tilde t_1}^2)
\Bigr \rbrace/m_t^2 \biggr ],
\label{mh-sq-primer}
\end{eqnarray}
where $c_{\tilde t}$ and $s_{\tilde t}$ are the cosine and sine of the stop
mixing angle.

To compensate for the tree level deficit at lower values of $\tan(\beta)$ we need sufficiently large stop masses which makes the Higgs exclusion contours more powerful, i.e. they exclude larger portions of the parameter space.

\bigskip
From Fig.~\ref{fig:GGM-sq-gl-plane} we see that in the GGM SUSY framework with not too high values of
$\tan(\beta)$, the current LHC searches for supersymmetry based on missing transverse momentum with jets and zero leptons
in the final state have only just started to be competitive with the Higgs exclusions from LEP.
In these scenarios we are not surprised that SUSY has not yet been discovered at LHC.

We note however, that the jets + MET searches exclude even decoupled squarks for sufficiently light
gluinos ($m_{\tilde{g}}\lesssim 500\,{\rm GeV}$). This is a region where the Higgs exclusion cannot compete.

The area of SUSY parameter space least probed and constrained by jets+MET SUSY searches is the area with a charged NLSP, in GGM notably the stau. This has already been noted in~\cite{Dolan:2011ie} and can be seen very clearly from Fig.~\ref{fig:GGM-sq-gl-plane}.
The Higgs exclusion holds regardless of the NLSP identity.
Therefore, it is crucial that the data from dedicated searches for charged massive particles started in~\cite{ATLAS-stau,CMS-stau} can be interpreted
in terms of general classes of SUSY models.
\bigskip

To what extent is the relative prominence of the Higgs exclusion in GGM different from other SUSY models, in particular from the
much studied example of the CMSSM? An important distinctive feature of all gauge mediation models in this context is the absence (or
more precisely the loop-level suppression) of the trilinear soft $A$-terms at the high scale. This is not the case in
gravity mediation models. In the CMSSM the $A$-coupling at the GUT scale is given by $A_0$ which is
a free parameter.

\begin{figure}
\begin{center}
  \includegraphics[width=10.5cm]{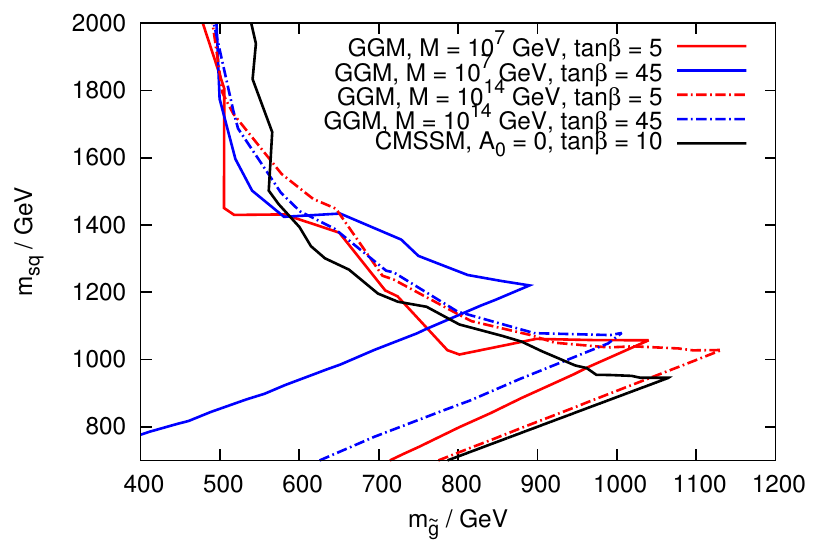}
\end{center}
\begin{center}
\caption{Jets + MET SUSY exclusion for the CMSSM, and for GGM for different values of $\tan (\beta)$ and the messenger scale $M$.
The diagonals delimiting the excluded areas in each case arise from one of the following two effects.
Either the NLSP (in gauge mediation models) changes from a neutralino to a stau, for which our implementation of the ATLAS jets + MET search has no sensitivity, or we simply reach a region in the gluino-squark mass plane which is theoretically inaccessible~\cite{Jaeckel:2011wp}.}
\label{fig:CMSSM-GGM-comb}
\end{center}
\end{figure}

We now switch to the CMSSM and compare the jets + MET exclusion with the LEP Higgs exclusion
at different $\tan(\beta)$ and $A_0$. The insensitivity of the former to these variables in the CMSSM was shown in \cite{Akula:2011zq}. Higgs bounds however depend on both. We illustrate this in Figure~\ref{fig:CMSSM-sq-gl-plane}. In the left panel we show $\tan(\beta)=5$, $10$ and $45$ for constant $A_0=0$, and
we reach similar conclusions as for the gauge mediation case discussed above.
In the right panel we vary $A_0$ from $-m_{0}$ to $+2m_{0}$ (where $m_{0}$ is the usual CMSSM scalar mass parameter), keeping $\tan(\beta) = 10$ fixed.
The variation of $A_0$ affects the value of the Higgs mass and the relevance of the Higgs exclusion contour.
The reason is that negative (positive) $A_{0}$ increases (decreases) stop mixing and thus the physical Higgs mass (see Eq.~\eqref{mh-sq-primer}), making the Higgs searches less (more) constraining.

In this sense, for the CMSSM and related models, the LEP Higgs bound provides a stronger constraint than jets + MET searches only in a part of the parameter space, characterised by vanishing or (moderately) positive values of $A_0$.
A common strategy for dealing with the additional parameter $A_{0}$ is,
instead of keeping it fixed, to optimise with respect to it.
In other words one looks for such values of $A_0$ where the SUSY is least constrained by the Higgs (and, if desired,
other experimental inputs~\cite{Buchmueller:2011sw,Buchmueller:2011aa}).
For GGM models, $A_0$ is not a free parameter and there is no simple way of reducing the dominance of the LEP Higgs exclusion contours (apart from dialing up $\tan(\beta)$).

We would like to note that even in models where $A_{0}$ is a free parameter, the direct searches provide additional sensitivity beyond the Higgs bounds
only for a certain range of $A_{0}$. As the overall scale probed by the collider increases this range grows as well. Therefore the size of the newly probed parameter space increases rapidly. Again SUSY searches are just taking off.

\bigskip
The observation that for not too high values of $\tan(\beta)\lesssim 10$, the available parameter space of general models of gauge mediation is not
yet significantly constrained by the current MET SUSY searches at the LHC is one of the main points of this paper. The ongoing searches of supersymmetry
at the LHC are entering the crucial discovery phase, but they are far from announcing the demise of supersymmetry.

\bigskip

The details of our analysis leading to the implementation and interpretation of the LHC SUSY searches for GGM models
as well as the validation of this implementation for the known CMSSM case will be presented in the following section.
There we also explain how the Higgs bounds were derived.
Before moving on, however, we note that the jets + MET exclusion shows little model dependence provided the neutralino is stable on detector scales.
In Figure~\ref{fig:CMSSM-GGM-comb} we show this exclusion in terms of physical squark and gluino masses both
for the CMSSM and for GGM with low and high values of $\tan(\beta)$ and of the messenger scale $M$.
As one can clearly see the variations in the relative positions of the
contours are not dramatic\footnote{This agrees with the observation~\cite{Desai:2011th} that the slepton mass has little effect on the exclusion.}. The position of the turning point where stau becomes lighter than the neutralino and the resulting slope
do however depend on the model.

\section{Details of the Analysis}

\subsection{Implementation and validation of the ATLAS jets + MET SUSY search}

Our approach in implementing and validating the latest $1.04\,{\rm fb}^{-1}$ ATLAS MET search constraints for a general BSM model
follows the general strategy laid out in Ref.~\cite{Dolan:2011ie} and
uses a combination of
\textsf{Herwig++}, \textsf{RIVET} and \textsf{Prospino}.

\begin{figure}
\vskip -1cm
\subfigure{
  \includegraphics[width=7.4cm]{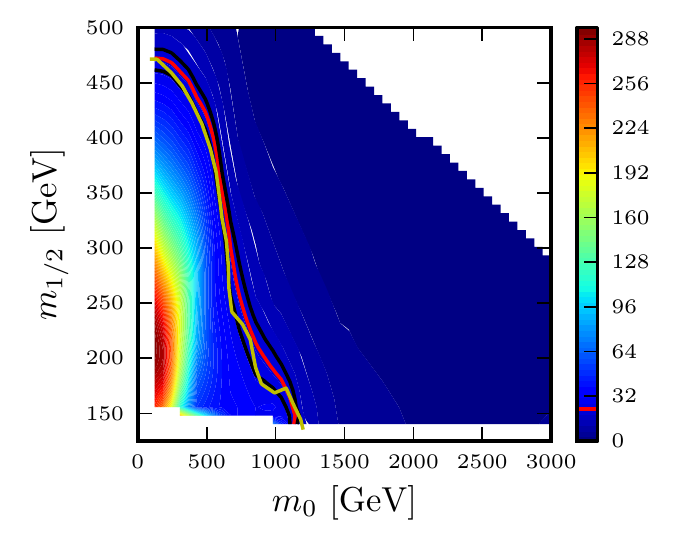}
  \hspace*{0.3cm}
  \includegraphics[width=7.4cm]{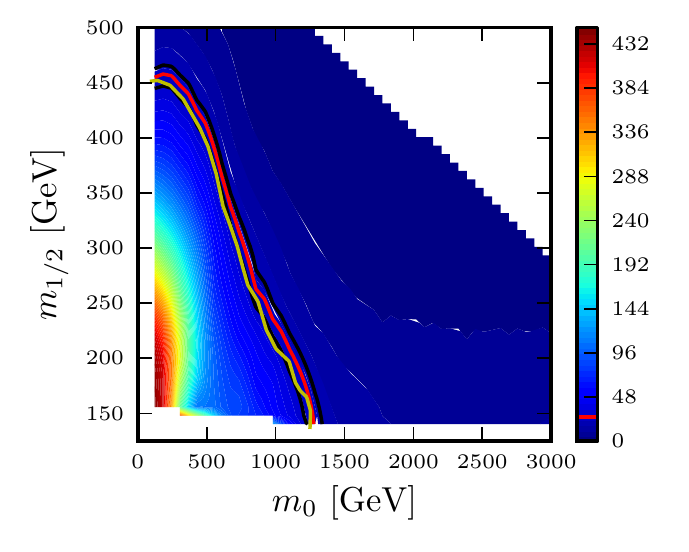}
  }
  \subfigure{
  \includegraphics[width=7.4cm]{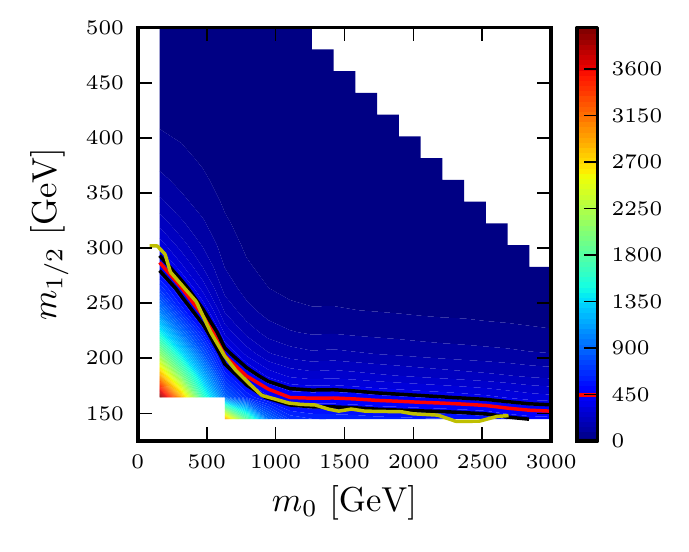}
  \hspace*{0.3cm}
  \includegraphics[width=7.4cm]{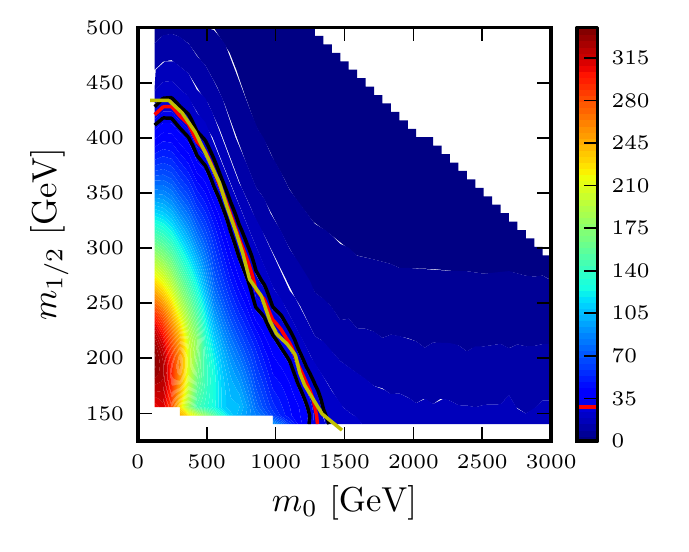}
  }
   \subfigure{
  \includegraphics[width=7.4cm]{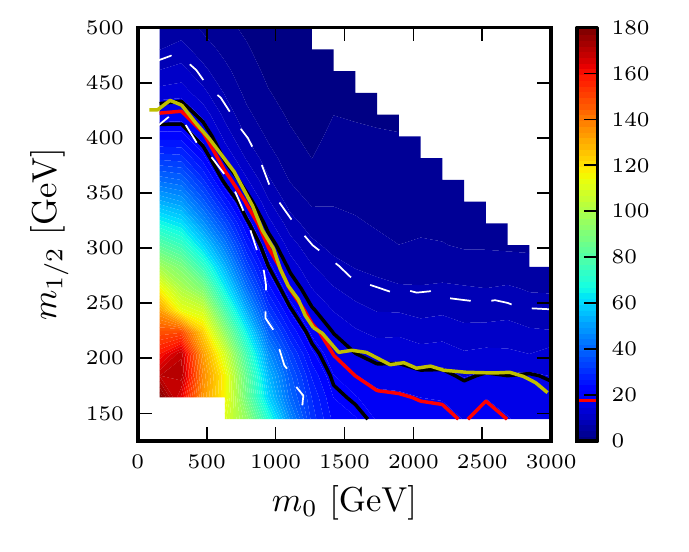}
  }
  \vspace*{-0.7cm}
\begin{center}
\caption{ 
Validation of our signal simulation in the CMSSM against the ATLAS limits obtained in \cite{ATLAS-0lept} in the five signal regions
defined in
their Table~2:
$\ge$~2-jets, $\ge$~3-jets, $\ge$~4-jets with $m_{\rm eff}> 500$~GeV, $\ge$~4-jets with $m_{\rm eff}> 1000$~GeV and ``High Mass'' counting
from left to right and top to bottom. We use an efficiency factor as described in the text.
The red curves denote the exclusion limits obtained from our simulation; its scale variation (changing the
renormalization/factorization scale by $2^{\pm1}$) is indicated by the black curves. The ATLAS exclusion is shown in yellow.
In the last panel -- the High Mass region -- we also show the $\pm 1 \sigma$ contours for the ATLAS expected exclusion (thin dashed white).
The wide splitting of these two contours shows the difficulty in predicting this contour's position precisely.
The colour indicates the cross section in ${\rm fb}$ in each of the signal regions. The red line in the colour bar shows the quoted limit.}
\label{fig:validate}
\end{center}
\end{figure}

A Monte Carlo event generator, in our case \textsf{Herwig++ 2.5.1}~\cite{Bahr:2008pv,Gieseke:2011na}, is used to compute
signal rates from any given BSM model, in our case GGM (or CMSSM for validation) with mass spectra calculated by \textsf{SoftSUSY 3.1.6}~\cite{Allanach:2001kg}.
Our implementation\footnote{It is publicly available as part of \textsf{RIVET} package: analysis {\it ATLAS\_2011\_S9212183}.} of the experimental event selection used by ATLAS in \cite{ATLAS-0lept}, defining five different signal regions, was done in the \textsf{RIVET 1.5.2}~\cite{Buckley:2010ar} analysis
framework.
Since the matrix elements implemented
in general purpose Monte Carlo event generators are only accurate to leading order
in perturbative QCD, we supplement the acceptance
calculated with \textsf{Herwig++} with an NLO cross section calculated with \textsf{Prospino 2.1}~\cite{Beenakker:1996ch,Beenakker:1999xh,Spira:2002rd,Plehn:2004rp}.
More precisely, we use \textsf{Herwig++} to simulate three sets of
supersymmetric particle production processes for
each point in SUSY parameter space:
a) squark and gluino production, b) the production of an electroweak gaugino in association with
a squark or gluino and c) the production of slepton and electroweak gaugino pairs. The fraction of events passing the experimental
cuts in each channel was then used together with the cross section calculated using
\textsf{Prospino} to obtain the number of signal events passing the cuts for each of the five signal regions.
The maximum allowed number of signal events in each signal region is provided by ATLAS~\cite{ATLAS-0lept}.
Comparing our numbers of events calculated in a given SUSY model point with this number decides whether the model point is allowed or excluded.

To validate our implementation we simulate the same CMSSM plane in the five signal regions presented by ATLAS in Ref.~\cite{ATLAS-0lept}
and compare our exclusion with theirs.
As can be seen in Fig.~\ref{fig:validate} we obtain excellent agreement between our exclusion contours and theirs. Having implemented the same
kinematic cuts and selections as ATLAS this consistency means that the detector efficiencies are relatively close to one in the relevant
kinematic regions. This makes us confident that our analysis, which does not use a detector simulation, can be applied to the GGM case.
As discussed in Ref.~\cite{ATLAS-0lept} an efficiency factor is needed to account for loss of jet energy. We have included
a factor of 0.85 for the four of five signal regions with softer cuts on the jet $p_{T}$ (i.e. excluding the High Mass region\footnote{For the High Mass region our agreement with the ATLAS exclusion contour is better without such
an efficiency factor. As can be seen from Fig.~\ref{fig:validate} the position of this contour has large uncertainties.}) to account for this and give a minor improvement.

\subsection{Higgs Bounds}

To check exclusion by Higgs-based searches, we used the superpartner spectrum computed by \textsf{SoftSUSY}
with FeynHiggs 2.8.5~\cite{Frank:2006yh,Degrassi:2002fi,Heinemeyer:1998np,Heinemeyer:1998yj}
to accurately calculate further details of the Higgs sector, in particular decay cross-sections.
This information was passed to {\tt HiggsBounds 3.5.0beta}~\cite{Bechtle:2008jh,Bechtle:2011sb} to compare with current
experimental limits from LEP, the Tevatron and the LHC (version 3.5.0 includes data from the EPS-HEP-2011,
LeptonPhoton-2011, and SUSY-2011 conferences). {\tt HiggsBounds} returns $\sigma / \sigma_{\text{limit}}$ for the search channel
with the highest statistical sensitivity: a ratio greater than $1$ indicates $95\%$ confidence-level exclusion. This approach has the rigour of Monte Carlo based exclusion -- in comparing cross-sections to their limits rather than using model-dependent mass limits -- whilst only taking a time comparable to the generation of the mass spectrum itself. In our case we see only the LEP lower bound, $m_{h}\gtrsim 114.4\,{\rm GeV}$, for an SM-like Higgs. However for more exotic SUSY scenarios, and with more data collected by the LHC and included in  {\tt HiggsBounds}, this may not be the case. (A user-friendly
python script for linking \textsf{SoftSUSY} with FeynHiggs with {\tt HiggsBounds} is
available at \href{http://www.ippp.dur.ac.uk/~SUSY}{\bf http://www.ippp.dur.ac.uk/$\sim$SUSY}.)

\subsection{Implementation of the ATLAS jets + MET SUSY search in GGM}\label{ggm}

\begin{figure}
\begin{center}
  \includegraphics[width=7.4cm]{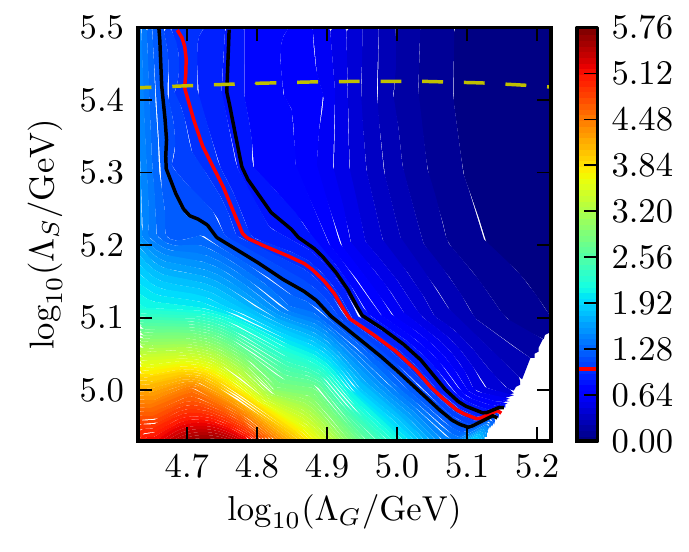}
  \hspace*{0.3cm}
  \includegraphics[width=7.4cm]{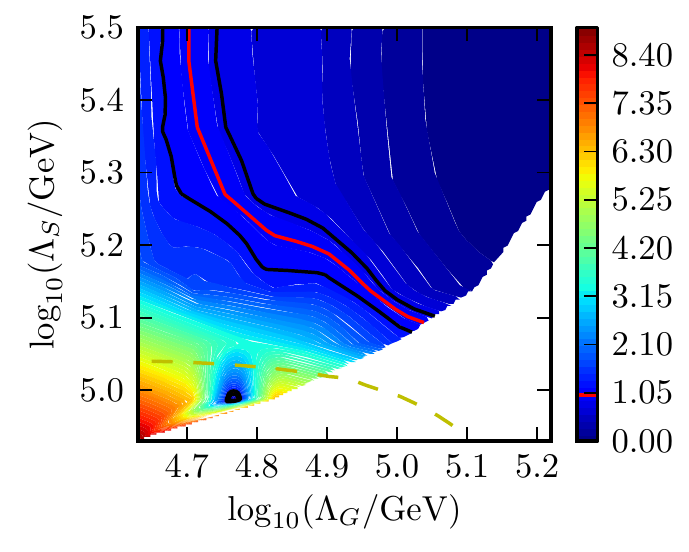}
  \includegraphics[width=7.4cm]{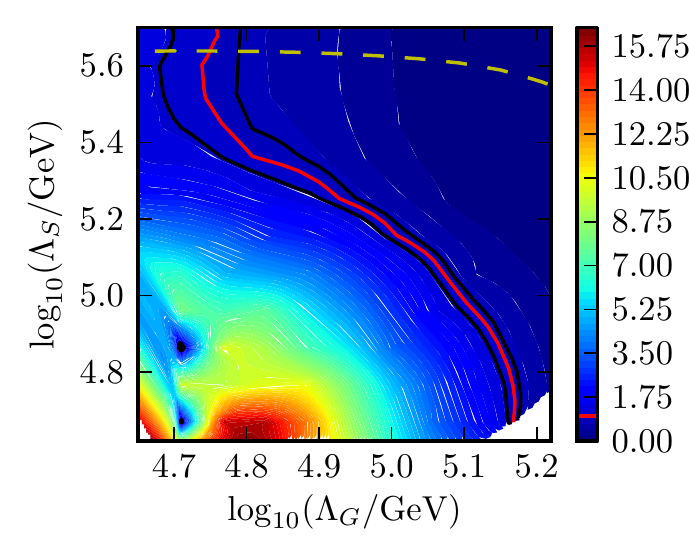}
  \hspace*{0.3cm}
  \includegraphics[width=7.4cm]{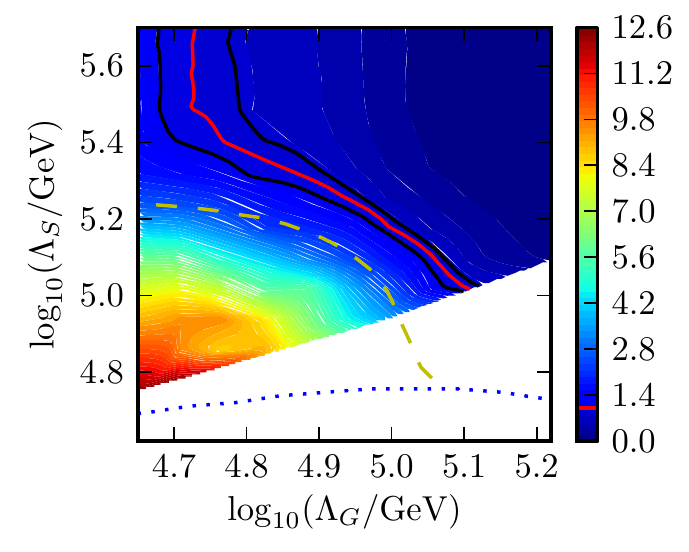}
\end{center}
\begin{center}
\caption{Exclusion contours for General Gauge Mediation from the ATLAS $1.04\,{\rm fb}^{-1}$ jets + MET SUSY search
and from the LEP Higgs bound, with a messenger scale $M=10^7$ GeV (top panels) and $M=10^{14}$ GeV (bottoms panels).
The left panels have $\tan (\beta) = 5$ and the right panels $\tan (\beta) = 45$.
The red line shows implementation of the ATLAS data for GGM; the black lines denote the scale variation. The yellow dashed line is the Higgs exclusion. White regions are stau NLSP. The blue dotted line in the bottom right panel delineates a tachyonic stau region.
The colour throughout each plot denotes $\sigma / \sigma_{\mathrm{limit}}$ for the most constraining region.}
\label{fig:GGM-LS-LG-7}
\end{center}
\end{figure}

In General Gauge Mediation~\cite{Meade:2008wd} the soft supersymmetry breaking gaugino masses at the messenger scale $M$ are given by
\begin{equation}
\label{gauginosoft}
M_{\tilde{\lambda}_i}(M) =\, k_i \,\frac{\alpha_i(M)}{4\pi}\,\Lambda_{G,i}
\end{equation}
where $k_i = (5/3,1,1)$, $k_i\alpha_i$ (no sum)
are equal at the GUT scale and $\alpha_i$ are the gauge coupling constants.
Similarly, the scalar mass squareds are
\begin{equation}
\label{scalarsoft}
m_{\tilde{f}}^2 (M) =\, 2 \sum_{i=1}^3 C_i k_i \,\frac{\alpha_i^2(M)}{(4\pi)^2}\, \Lambda_{S,i}^2
\end{equation}
where the $C_i$ are the quadratic Casimir operators of the gauge groups.

The value of the messenger scale $M$, the value of $\tan(\beta)$ together with $\Lambda_G$ and $\Lambda_S$ appearing
in Eqs.~\eqref{gauginosoft}-\eqref{scalarsoft}
at  $M$ characterise a point in the GGM
parameter space~\cite{Jaeckel:2011ma}.

For clarity we show numerical results for the case where, as in~\cite{pureGGM1,pureGGM2}, we do not further split $\Lambda_G$ or $\Lambda_S$
into separate independent factors for each gauge group (this also ensures automatic gauge unification). At the end of this section we explain, however,
that our findings hold also for large regions of the full GGM parameter space with separate $\Lambda$ for each gauge group.
Furthermore, and in contrast to~\cite{pureGGM1,pureGGM2}, we do \emph{not} confine ourselves here to {\it pure GGM} models where
$\tan (\beta)$ is determined from the vanishing input value of the $B$ soft term.
Here $\tan (\beta)$ is treated as a free parameter.

Before we proceed let us briefly comment on the nature of the NLSP in gauge mediation\footnote{A detailed study for the simpler case of pure GGM
can be found in Sect.~3 of Ref.~\cite{pureGGM2}.}.
As is well known the LSP in gauge mediation is the gravitino. The collider phenomenology then depends crucially on the nature of the NLSP and its lifetime. In general gauge mediation there is a wide range of options for what can be the NLSP. However the jets + MET search as implemented in this
paper is most sensitive to long-lived neutralino NLSPs. A long lived neutralino naturally leads to significant amounts of missing energy.
On the contrary a non-neutral NLSP like, e.g., a stau would leave a track and therefore wouldn't be counted as missing energy.
Similarly if the neutralino decays inside the detector into, say, a photon and a gravitino, then less missing energy is generated (and the event may also be vetoed due to the presence of photons).
For these reasons the zero-lepton jets + MET searches are primarily aimed at and are most sensitive to long-lived neutralinos.
This type of NLSP arises in large parts of the GGM parameter space. Specifically, the NLSP decay length is approximately given by,
\begin{equation}
L_{\rm NLSP}\sim \frac{1}{k^2_{G}}\left(\frac{100\,{\rm GeV}}{m_{\rm NLSP}}\right)^5 \left(\frac{\sqrt{\Lambda_{G} M}}{100\,{\rm TeV}}\right)^4\, 0.1\, {\rm mm},
\end{equation}
where $k_{G}:=1/C_{\rm grav}$ quantifies the coupling of messengers to the SUSY breaking sector which can take values of order 1 but is often much smaller. For $\Lambda_{G}\sim 10^{5}\,{\rm GeV}$, already moderate messenger scales of the order of $10^{7}\,{\rm GeV}$, lead to $L_{\rm decay}\gtrsim 10$~m and a decay outside of the detector. When $k_{G}\ll 1$ even smaller values of the $M$ may suffice.
This addresses the longevity of the NLSP.
Turning now to the NLSP species we find in the simple setup with a single $\Lambda_{G}$ and a single $\Lambda_{S}$ that roughly half of the parameter space is neutralino (the rest is mostly stau). The NLSP content of the current model is very similar (especially at high values of $\tan\beta$)
to that of pure GGM which is shown in Figure~5 of Ref.~\cite{pureGGM2}.

\bigskip
Figure~\ref{fig:GGM-LS-LG-7} shows our results for the jets + MET exclusion of GGM in terms of the original $\Lambda_G$ and $\Lambda_S$ model parameters. The upper panels show a messenger scale $M=10^7$ GeV, the lower panels $M=10^{14}$ GeV. The left panels have $\tan(\beta)=5$ and the right panels $\tan(\beta)=45$.

The relative importance of Higgs (yellow dashed line) vs direct SUSY exclusion contours in GGM for various values of $\tan(\beta)$ (and $M$) can be directly read off
these $\Lambda_{G}$-$\Lambda_{S}$ planes.
The projection of Fig.~\ref{fig:GGM-LS-LG-7} to the physical $m_{\tilde{g}}$-$m_{sq}$ plane is Fig.~\ref{fig:GGM-sq-gl-plane}.

\bigskip
Let us conclude with a brief discussion of what happens if we allow for six different $\Lambda$s to span the entire GGM parameter space.
We need to address what happens in this case for both the direct SUSY search and Higgs exclusion contours.
Jet + MET SUSY searches depend crucially on three parameters, the mass of the gluinos, the first generation squark mass and the neutralino mass. The first two are taken into account by presenting our results in Fig.~\ref{fig:GGM-sq-gl-plane} on the gluino-squark mass plane. The gluino mass chiefly depends on the parameter $\Lambda_{G,3}$. For direct searches the most relevant squark mass is the lightest one. This squark mass squared is determined by a certain linear combination of the $\Lambda^{2}_{S,i}$ and the $\Lambda^2_{G,i}$ through the RG evolution.
As long as the neutralino remains the NLSP and its mass is not fine-tuned to be very close to the squark and gluino masses, the exclusion
limits from direct searches in the gluino-squark mass plane are not affected by a splitting of $\Lambda$s.
On the other hand if the splitting of $\Lambda$s changes NLSP species jets + MET searches lose sensitivity, as explained earlier.

Now the Higgs exclusion contour chiefly depends on the stop masses. Since gauge mediation does not distinguish between the generations
this mass is directly linked to the mass of the first generation squarks. Unless extreme splittings are introduced this is essentially
the same squark mass parameter that governs the direct searches. In addition there is a higher order dependence of the Higgs mass on the gluino mass, which is our second parameter. In summary the same two parameters, squark and gluino mass, are the most relevant ones for both, the direct SUSY searches and the Higss constraints. Showing the results in the plane spanned by these two parameters gives qualitatively the same picture independent of the particular combination of the original high energy parameters, $\Lambda$.

\bigskip
\subsection*{Note added}
To bring the Higgs mass in our GGM model up into the region $m_{h_{0}}\sim 125\,{\rm GeV}$, hinted at by the recent LHC results~\cite{ATLASHiggs,CMSHiggs}, requires
fairly high superpartner masses. If we want to keep gluino masses as low as possible ($\sim 500\,{\rm GeV}$, cf. Fig.~\ref{fig:GGM-sq-gl-plane}),
we find that the squark masses must be in the region of $\gtrsim 10\,{\rm TeV}$.

\section*{Acknowledgements}
We thank Matt Dolan, Christoph Englert and Michael Spannowsky for useful discussions and Karina Williams for help
with {\tt HiggsBounds} and FeynHiggs.
This work was supported by STFC.

\newpage

\providecommand{\href}[2]{#2}\begingroup\raggedright
\endgroup
\end{document}